\documentclass[twocolumn,aps,prd,10pt,superscriptaddress,longbibliography]{revtex4-1}
\usepackage{amsmath}
\usepackage{bm}
\usepackage{graphicx}
\usepackage{multirow}
\usepackage[usenames,dvipsnames,svgnames]{xcolor}
\usepackage{hyperref}
\hypersetup{
pdfnewwindow=true,      
colorlinks=true,        
linkcolor=Blue,         
citecolor=Blue,         
filecolor=Blue,         
urlcolor=Blue           
}

\begin{document}

\title{Neuroevolution machine learning potentials: Combining high accuracy and low cost in atomistic simulations and application to heat transport}

\author{Zheyong Fan}
\email{brucenju@gmail.com}
\affiliation{College of Physical Science and Technology, Bohai University, Jinzhou, P. R. China}
\affiliation{MSP group, QTF Centre of Excellence, Department of Applied Physics, Aalto University, FI-00076 Aalto, Espoo, Finland}
\author{Zezhu Zeng}
\affiliation{Department of Mechanical Engineering, The University of Hong Kong, Pokfulam Road, Hong Kong SAR, China}
\author{Cunzhi Zhang}
\affiliation{Pritzker School of Molecular Engineering, The University of Chicago, Chicago, Illinois 60637, United States}
\author{Yanzhou Wang}
\affiliation{Beijing Advanced Innovation Center for Materials Genome Engineering, Department of Physics, University of Science and Technology Beijing, Beijing 100083, China}
\affiliation{MSP group, QTF Centre of Excellence, Department of Applied Physics, Aalto University, FI-00076 Aalto, Espoo, Finland}
\author{Keke Song}
\affiliation{Beijing Advanced Innovation Center for Materials Genome Engineering, Department of Physics, University of Science and Technology Beijing, Beijing 100083, China}
\author{Haikuan Dong}
\affiliation{College of Physical Science and Technology, Bohai University, Jinzhou, P. R. China}
\affiliation{MSP group, QTF Centre of Excellence, Department of Applied Physics, Aalto University, FI-00076 Aalto, Espoo, Finland}
\affiliation{Beijing Advanced Innovation Center for Materials Genome Engineering, Corrosion and Protection Center, University of Science and Technology Beijing, Beijing 100083, China}
\author{Yue Chen}
\affiliation{Department of Mechanical Engineering, The University of Hong Kong, Pokfulam Road, Hong Kong SAR, China}
\author{Tapio Ala-Nissila}
\affiliation{MSP group, QTF Centre of Excellence, Department of Applied Physics, Aalto University, FI-00076 Aalto, Espoo, Finland}
\affiliation{Interdisciplinary Centre for Mathematical Modelling, Department of Mathematical Sciences, Loughborough University, Loughborough, Leicestershire LE11 3TU, UK}

\date{\today}

\begin{abstract}
We develop a neuroevolution-potential (NEP) framework for generating neural network based machine-learning potentials.
They are trained using an evolutionary strategy for performing large-scale molecular dynamics (MD) simulations. A descriptor of the atomic environment is constructed based on Chebyshev and Legendre polynomials. The method is implemented in graphic processing units within the open-source \textsc{gpumd} package, which can attain a computational speed over $10^7$ atom-step per second using one Nvidia Tesla V100. Furthermore, per-atom heat current is available in NEP, which paves the way for efficient and accurate MD simulations of heat transport in materials with strong phonon anharmonicity or spatial disorder, which usually cannot be accurately treated either with traditional empirical potentials or with perturbative methods.
\end{abstract}

\maketitle

\section{Introduction}

Classical interatomic potentials play a crucial role in atomistic simulations, in particular in molecular dynamics (MD) simulations, where various static and dynamical materials properties can be efficiently computed. Machine-learning (ML) potentials (or force fields) \cite{behler2016jcp,Deringer2019am,Mueller2020jcp,Mishin2021am,Unke2021cr}, i.e., interatomic potentials constructed based on a ML model, have been demonstrated to be able to achieve an accuracy comparable to their quantum-mechanical training data while reducing the computation time to a small fraction of the corresponding quantum-mechanical calculations.

Various ML models have been used to construct ML potentials, including e.g., artificial neural networks  \cite{behler2007prl}, Gaussian regression \cite{bartok2010prl}, and linear regression \cite{Thompson2014jcp}. For any ML model, there are many fitting parameters that need to be determined by training the model against quantum-mechanical data. The large number of fitting parameters is the very foundation for the superior interpolation capability of ML potentials as compared to conventional empirical potentials that only have a few to a few tens of fitting parameters. However, finding an optimized set of parameters is a nontrivial task. The conventional training method in neural network potentials is based on gradient descent, which could be trapped into a local minimum of the loss function of a ML model, leading to a sub-optimal solution.

An alternative training method for ML models is based on evolutionary algorithms, such as genetic algorithms, genetic programming, evolutionary programming, and evolutionary strategy. This global-searching approach combined with neural networks is known as neuroevolution, and it has long been applied to evolve neural networks \cite{yao1999}. It has been greatly improved by state-of-the-art evolutionary algorithms such as the natural evolution strategies \cite{wierstra2014jmlr}. A variant called separable natural evolution strategies \cite{Schaul2011} has a computational complexity linear in the number of fitting parameters and is well suited to evolve large-scale neural networks. 

In this work, we develop a framework called neuroevolution potentials (NEPs) for generating neural network based ML potentials trained using the separable natural evolution strategy \cite{Schaul2011}. Although evolutionary algorithms are less likely to be trapped into a local minimum, these population based methods require evaluating the loss function multiple times within one step (also called one generation) and are usually more computationally demanding than gradient-descent-based algorithms. To speed up the calculations, we realize an efficient graphics processing units (GPU) implementation of the calculations within the open-source \textsc{gpumd} package \cite{fan2013cpc,fan2017cpc,gpumd}. The GPU implementation both speeds up the training process and makes MD simulations significantly faster than the current implementations of ML potentials. We demonstrate the efficiency and accuracy of NEP by comparing \textsc{gpumd} with some popular implementations of ML potentials, including the \textsc{quip} package \cite{quip} that implements the Gaussian approximation potential (GAP) \cite{bartok2010prl}, the \textsc{mlip} package \cite{Novikov2021} that implements the moment tensor potential (MTP) \cite{Shapeev2016}, and the DeePMD-kit package \cite{wang2018cpc} that implements the deep potential (DP) \cite{zhang2018prl,zhang2018endtoend}. In particular, here we focus on demonstrating the applicability of NEP in heat transport simulations.

The remainder of this paper is organized as follows. In Sec. \ref{section:general} and Sec. \ref{section:virial-heat}, we review the formulations of the per-atom force, virial and heat current for general many-body potentials \cite{fan2015prb,Gabourie2021}. In Sec. \ref{section:descriptors}, we present the descriptor used in NEP, which is a mapping from a set of relative coordinates to a set of functions with required symmetries. In Sec. \ref{section:potential}, the neural network connecting the descriptor and the site energy of an atom is discussed. Section \ref{section:train} presents detailed algorithms for training NEP using a natural evolution strategy. Section \ref{section:cuda} discusses the general strategies in our GPU implementation. In Sec. \ref{section:performance}, we validate and benchmark NEP by comparing its performance with some other implementations of ML potentials. In Sec. \ref{section:heat}, we demonstrate the applicability of NEP in heat transport simulations with selected case studies. Section \ref{section:summary} summarizes and concludes.

\section{Theory}

Machine-learning potentials are usually many-body potentials. To make an efficient GPU implementation of a many-body potential, one must first derive explicit expressions of various per-atom quantities \cite{fan2017cpc}. We start by reviewing some formulations as derived in Refs. \onlinecite{fan2015prb,Gabourie2021}.

\subsection{General many-body potential and partial forces\label{section:general}}

For a general many-body potential, the total potential energy $U$ of a system can be written as
\begin{equation}
    U = \sum_i U_i,
\end{equation}
where the site energy $U_i$ of atom $i$ is 
\begin{equation}
    U_i = U_i \left(
    \{\bm{r}_{ij}\}
    \right).
\end{equation}
In this paper, we define $\bm{r}_{ij}$ as the relative position from atom $i$ to atom $j$, that is,
\begin{equation}
\bm{r}_{ij} \equiv \bm{r}_{j} - \bm{r}_{i}.
\end{equation}
The Cartesian components for this vector will be denoted as $x_{ij}$, $y_{ij}$, and $z_{ij}$. Therefore, $\{\bm{r}_{ij}\}$ denotes the collection of relative positions from the central atom $i$ to all the other atoms $j$. Usually, a finite cutoff distance $r_{\rm c}$ is adopted such that only the atoms $j$ with a distance $r_{ij}$ to $i$ that is smaller than $r_{\rm c}$ are considered in the collection.

Starting from the potential energy above, a general force expression which respects the weak form of Newton's third law has been derived as \cite{fan2015prb}
\begin{equation}
    \bm{F}_{i} = \sum_{j\neq i} \bm{F}_{ij};
\end{equation}
\begin{equation}
    \bm{F}_{ij} = - \bm{F}_{ji} = \frac{\partial U_i}{\partial \bm{r}_{ij}} - \frac{\partial U_j}{\partial \bm{r}_{ji}}.
\end{equation}
Here, $\bm{F}_{ij}$ can be understood as the force acting on atom $i$ from atom $j$, possibly influenced by other atoms. The partial derivative $\partial/ \partial \bm{r}_{ij}$ should be understood as a vector with the components $\partial/\partial x_{ij}$, $\partial/\partial y_{ij}$, and $\partial/\partial z_{ij}$. We note that although $\bm{F}_{ij}$ respects the weak form of Newton's third law, $\bm{F}_{ij}=-\bm{F}_{ji}$, it does not respect the strong form of Newton's third law, $\bm{F}_{ij}\propto \bm{r}_{ij}$, but all the formulations in Ref. \onlinecite{fan2015prb} only require the weak form.

Since $\partial U_j/\partial \bm{r}_{ji}$ can be obtained from $\partial U_i/ \partial \bm{r}_{ij}$ by an exchange of indices, $i\leftrightarrow j$, in practical implementation, we only need to calculate and store all the $\partial U_i/ \partial \bm{r}_{ij}$. Due to the importance of $\{\partial U_i/ \partial \bm{r}_{ij}\}$, we call them ``partial forces''. We will derive explicit expressions of these partial forces in Sec. \ref{section:potential}.

\subsection{Virial stress, heat current and thermal conductivity\label{section:virial-heat}}

Starting from the force expression, one can derive the per-atom virial tensor, which plays a crucial role in MD simulations. An expression has been derived in Ref. \onlinecite{fan2015prb}, but in Ref. \onlinecite{Gabourie2021}, it has been reformulated  into a more convenient form:
\begin{equation}
    \mathbf{W}_{i} = \sum_{j\neq i} \bm{r}_{ij} \otimes \frac{\partial U_j}{\partial \bm{r}_{ji}}.
\label{equation:Wi}
\end{equation}
Here, $\otimes$ represents the tensor product between two vectors. Using this per-atom virial expression, the per-atom heat current derived in Ref. \onlinecite{fan2015prb} can be conveniently expressed as
\begin{equation}
\bm{J}_{i} = \mathbf{W}_{i} \cdot \bm{v}_i,
\end{equation}
where $\bm{v}_i$ is the velocity of atom $i$. The per-atom virial in Eq. (\ref{equation:Wi}) is generally not a symmetric tensor and one must, in heat transport applications, use the full $3\times 3$ tensor instead of six components of it only. The total virial in a system, $\sum_i \mathbf{W}_{i}$, on the other hand, is a symmetric tensor with six independent components only. The validity of our heat current expression has been extensively documented \cite{Gill-Comeau2015prb,fan2017prb,xu2018msmse}, while that implemented in the widely-used \textsc{lammps} package \cite{plimpton1995jcp} has been shown \cite{fan2015prb,Gill-Comeau2015prb,Surblys2019pre,Boone2019jctc} to be erroneous for general many-body potentials.

The total heat current $\bm{J}$ is the sum of the per-atom heat currents $\bm{J}=\sum_i \bm{J}_i$. In the homogeneous nonequilibrium MD (HNEMD) method \cite{Fan2019prb}, the lattice thermal conductivity tensor $\kappa_{\mu\nu}$ can be computed from the following expression,
\begin{equation}
\frac{\langle J_{\mu}(t)\rangle_{\rm ne}}{TV} = \sum_{\nu} \kappa_{\mu\nu}  F^{\nu}_{\rm e},
\end{equation}
where $\bm{F}_{\rm e}$ is the driving force parameter and $\langle J_{\mu}(t)\rangle_{\rm ne}$ is nonequilibrium ensemble average of the heat current. The driving force parameter will induce the following external force on atom $i$ \cite{Fan2019prb,Gabourie2021}:
\begin{equation}
\bm{F}_{i}^{\rm ext}
= \bm{F}_{\rm e} \cdot \mathbf{W}_i
\end{equation}
for solid systems. The spectral lattice thermal conductivity $\kappa_{\mu\nu}(\omega)$ as a function of the angular frequency $\omega$ can be computed from the following relation \cite{Fan2019prb,Gabourie2021}:
\begin{equation}
\frac{2}{TV} \int_{-\infty}^{\infty} K_{\mu}(t) e^{i\omega t}dt = \sum_{\nu} \kappa_{\mu\nu}(\omega)  F^{\nu}_{\rm e},
\end{equation}
where 
\begin{equation}
\bm{K}(t) = \sum_i\langle \mathbf{W}_i(0) \cdot \bm{v}_i(t)\rangle_{\rm ne}  
\end{equation}
is the virial-velocity correlation function.

In summary, as long as the partial forces $\partial U_i/\partial \bm{r}_{ij}$ are derived and implemented, one can conveniently calculate the per-atom virial and heat currents, which can be used to realize the constant stress ensembles and detailed heat transport simulations.

\subsection{From coordinates to a descriptor vector \label{section:descriptors}}

In a ML potential, the site potential $U_i$ is not directly modelled as a function of the relative coordinates $\{\bm{r}_{ij}\}$, but as a function of a high-dimensional descriptor vector, whose components are invariant with respect to spatial translation, three-dimensional rotation and inversion, and permutation of atoms with the same species \cite{bartok2013prb}. Many descriptors have been proposed, including, e.g., Behler's symmetry functions \cite{behler2011jcp}, the smooth overlap of atomic positions (SOAP) \cite{bartok2013prb}, the bispectrum \cite{bartok2010prl}, the Coulomb matrix \cite{Rupp2012prl}, the moment tensor \cite{Shapeev2016}, the atomic cluster expansions \cite{drautz2019prb}, the embedded atom descriptor \cite{zhang2019jpcl}, the Gaussian moments \cite{Zaverkin2020jctc}, and the atomic permutationally invariant polynomials \cite{oord2020}. There are libraries implementing various descriptors \cite{Khorshidi2016cpc,Lauri2020cpc,Yanxon2021}.

\subsubsection{Single-component systems}

The descriptor we use in NEP is motivated by both Behler's symmetry functions \cite{behler2011jcp} and an optimized version of SOAP \cite{Caro2019prb}. For a central atom $i$ in a single-component system, we define a set of radial descriptor components ($n\geq 0$)
\begin{equation}
\label{equation:qin}
q^i_{n}
= \sum_{j\neq i} g_n(r_{ij}),
\end{equation}
and a set of angular descriptor components ($n\geq 0$ and $l\geq 1$)
\begin{equation}
q^i_{nl} 
= \sum_{j\neq i}\sum_{k\neq i} g_n(r_{ij}) g_n(r_{ik})
P_l(\cos\theta_{ijk}),
\label{equation:qinl}
\end{equation}
where $P_l(\cos\theta_{ijk})$ is the Legendre polynomial of order $l$, $\theta_{ijk}$ being the angle formed by the $ij$ and $ik$ bonds. The functions $g_n(r_{ij})$ are radial functions and we choose to express them as the first-kind Chebyshev polynomials of the variable $x \equiv 2(r_{ij}/r_{\rm c}-1)^2-1$:
\begin{equation}
\label{equation:g_n}
g_n(r_{ij}) = \frac{T_n(x)+1}{2} f_{\rm c}(r_{ij}).
\end{equation}
The variable $x$ is defined to have values from $-1$ to $1$.
Recurrence relations for evaluating the Chebyshev polynomials and their derivatives used here are presented in Appendix \ref{appendix:chebyshev}. Empirically, we found it beneficial to make $g_n(r_{ij})$ positive definite, as in Eq. (\ref{equation:g_n}). A similar expression has been used in the atomic cluster expansions approach \cite{drautz2019prb}.

\begin{figure}[htb]
\begin{center}
\includegraphics[width=\columnwidth]{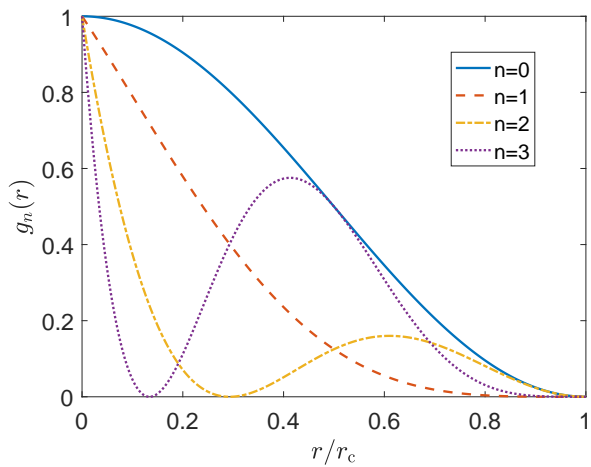}
\caption{The radial function $g_n(r)$ as a function of the reduced distance $r/r_{\rm c}$ for $0\leq n \leq 3$.}
\label{figure:gn}
\end{center}
\end{figure}

The function $f_{\rm c}(r_{ij})$ is a cutoff function defined as
\begin{equation}
   f_{\rm c}(r_{ij}) 
   = \frac{1}{2}\left(
   1 + \cos\left( \pi \frac{r_{ij}}{r_{\rm c}} \right) 
   \right)
\end{equation}
for $r\leq r_{\rm c}$, and  $f_{\rm c}(r_{ij}) =0$ for $r > r_{\rm c}$,
following the definition in the Behler-Parrinello neural network potential \cite{behler2007prl,behler2011jcp}. The first four lowest-order radial functions are shown in Fig. \ref{figure:gn}.

Note that radial functions are used both in the radial and angular components of the descriptor. The radial expansion in the radial components is up to a given order $n_{\rm max}^{\rm R}$, i.e., $n=0,1,\cdots, n_{\rm max}^{\rm R}$. For the angular components, the radial expansion is up to $n_{\rm max}^{\rm A}$, i.e., $n=0,1,\cdots, n_{\rm max}^{\rm A}$, and the angular expansion is up to $l_{\max}$, i.e., $l=1,\cdots, l_{\rm max}$. The full descriptor vector has a dimension of
\begin{equation}
\label{equation:N_des}
N_{\rm des} = \left(n_{\rm max}^{\rm R}+1\right) 
+ \left(n_{\rm max}^{\rm A}+1\right)l_{\rm max}. 
\end{equation}
Note that the cutoff radii for the radial and angular components are not necessarily the same, but can take different values denoted as $r_{\rm c}^{\rm R}$ and $r_{\rm c}^{\rm A}$, respectively. The radial components can be used to represent relatively long-ranged interactions (such as Coulomb and van der Waals interactions) and the angular components mainly account for intermediate-ranged interactions. 

If we change the factor  $g_{n}(r_{ij})g_{n}(r_{ik})$ in Eq. (\ref{equation:qinl}) to $g_{n}(r_{ij})g_{n'}(r_{ik})(2l+1)/4\pi$ and apply the addition theorem of spherical harmonics, we obtain a descriptor $q^i_{nn'l}$ similar to the SOAP descriptor in Ref. \onlinecite{Caro2019prb}:
\begin{equation}
q^i_{nn'l} = \sum_{m=-l}^{l} c^i_{nlm} \left(c^i_{n'lm}\right)^{\ast};
\end{equation}
\begin{equation}
c^i_{nlm} = \sum_{j\neq i} g_n(r_{ij})
Y_{lm}^{\ast} (\theta_{ij},\phi_{ij}),
\end{equation}
where $\theta_{ij}$ and $\phi_{ij}$ are the polar and azimuthal angles of the relative position $\bm{r}_{ij}$
in spherical coordinates. The major remaining difference between the angular components of our descriptor and the SOAP one in Ref. \onlinecite{Caro2019prb} is that the radial function in the latter also depends on $l$, and a different radial expansion scheme was adopted. The connections between different atom environment descriptors have been revealed recently \cite{Willatt2019jcp,drautz2019prb}.

\subsubsection{Multicomponent systems}

We have omitted the atom species in the discussion above. For a multicomponent system, a feasible method for constructing permutation-symmetric descriptors is to multiply the terms in Eq. (\ref{equation:qin}) by a weighting factor such as $Z_j$ and multiply the terms in Eq. (\ref{equation:qinl}) by a weighting factor such as $Z_jZ_k$  \cite{Gastegger2018jcp}, where $Z_j$ is the atomic number of atom $j$, although other weighting factors \cite{Artrith2017prb} can also be used.  This method has been adopted in the \verb"PyXtal_FF" package \cite{Yanxon2021} for all the descriptors implemented therein. Based on our definition of $g_n(r_{ij})$ in Eq. (\ref{equation:g_n}), this leads to a modification of the cutoff function: $f_{\rm c}(r_{ij}) \to f_{\rm c}(r_{ij}) Z_j$.

Here, we use the following modified definition:
\begin{equation}
    f_{\rm c}(r_{ij}) \to f_{\rm c}(r_{ij}) \sqrt{Z_i Z_j}.
\end{equation}
That is, we first change the factor $Z_j$ to $\sqrt{Z_j}$,
to make the relative weights less important in systems with very different atomic numbers, and then add the information of the central atom $\sqrt{Z_i}$ to
distinguish two configurations with the same environments but different central atom species, which are indistinguishable in the previous approaches \cite{Gastegger2018jcp,Artrith2017prb}. 

\subsection{From descriptor to site energy  \label{section:potential}}

We have stated above that in a ML potential, the site energy is taken as a function of the set of descriptor components $\{q^i_{nl}\}$:
\begin{equation}
    U_i = U_i(\{q^i_{nl}\}),
\end{equation}
which is a many-variable scalar function. Different ML models have been used to construct this many-variable function, including neural network \cite{behler2007prl}, Gaussian regression \cite{bartok2010prl}, and linear regression \cite{Thompson2014jcp}. In NEP, we choose the feedforward neural network (also called multilayer perceptron) as the ML model.

\subsubsection{The neural network model}

As in previous works \cite{behler2007prl,behler2011jcp}, the descriptor vector is taken as the input layer of the neural network and the site energy is taken as the output layer.  Between them, there can be one or more hidden layers of neurons (nodes). For simplicity, we assume a single hidden layer in the following presentation; generalization to more hidden layers is straightforward. 

The descriptor vector as the input layer is coupled to the (first) hidden layer. To facilitate the presentation, we combine the two labels $n$ and $l$ into a single one, $\nu=(nl)$, and write the descriptor vector for atom $i$ as $q^i_{\nu}$ ($1\leq \nu \leq N_{\rm des}$). The state of the hidden layer can also be represented as a vector, $x_{\mu}$ ($1\leq \mu \leq N_{\rm neu}$), where $N_{\rm neu}$ is the number of neurons in the hidden layer. The hidden layer state vector is obtained from the input vector by a combination of linear and nonlinear transforms:
\begin{equation}
\label{equation:nn1}
    x_{\mu} = \tanh\left(\sum_{\nu=1}^{N_{\rm des}} w^{(1)}_{\mu\nu} q^i_{\nu} - b^{(1)}_{\mu}\right),
\end{equation}
where $w^{(1)}_{\mu\nu}$ is the connection weight between the neurons $x_{\mu}$ and $q^i_{\nu}$, and $b^{(1)}_{\mu}$ is the bias for the neuron $x_{\mu}$. We choose the hyperbolic tangent function as the nonlinear transform, also called the activation function in the hidden layer, in agreement with previous works \cite{behler2011jcp,zhang2018prl}. The output layer state, which is the site energy, is calculated as a linear combination of the state vector of the hidden layer:
\begin{equation}
\label{equation:nn2}
U_i = \sum_{\mu=1}^{N_{\rm neu}}w^{(2)}_{\mu}x_{\mu} - b^{(2)},
\end{equation}
where $w^{(2)}_{\mu}$ is the connection weight between the neurons $U_i$ and $x_{\mu}$, and $b^{(2)}$ is the bias for the neuron $U_i$.

\subsubsection{An explicit expression for the partial force}

We can now present an explicit expression for the partial force defined in Sec. \ref{section:general} as
\begin{equation}
\frac{\partial U_i}{\partial \bm{r}_{ij}} 
= \sum_n\sum_l \frac{\partial U_i}{\partial q^i_{nl}} \frac{\partial q^i_{nl}}{\partial \bm{r}_{ij}}.
\end{equation}
The factor $\partial U_i/\partial q^i_{nl}$ can be calculated based on Eqs. (\ref{equation:nn1}) and (\ref{equation:nn2}). The factor $\partial q^i_{nl}/\partial \bm{r}_{ij}$ can be calculated based on our descriptor expressions. For the radial components, we have
\begin{equation}
\frac{\partial q^i_{n}}{\partial \bm{r}_{ij}} =  \frac{\partial g_n(r_{ij})}{\partial r_{ij}} \frac{\bm{r}_{ij}}{r_{ij}}.
\end{equation}
For the angular components, we have
\begin{align}
\frac{\partial q^i_{nl}}{\partial \bm{r}_{ij}} 
&= 2\sum_{k\neq i}  \frac{\partial g_n(r_{ij})}{\partial r_{ij}} g_n(r_{ik})  
\frac{\bm{r}_{ij}}{r_{ij}}  P_l(\cos\theta_{ijk}) \nonumber \\
&+ 2 \sum_{k\neq i}  g_n(r_{ij}) g_n(r_{ik})  
\frac{\partial P_l(\cos\theta_{ijk})}{\partial \cos\theta_{ijk}}
\nonumber \\
&\times \frac{1}{r_{ij}}
\left(\frac{\bm{r}_{ik}}{r_{ik}} - \frac{\bm{r}_{ij}}{r_{ij} }\cos\theta_{ijk}\right).
\end{align}

\subsection{Training the machine learning potential\label{section:train}}

\subsubsection{Defining a loss function}

The purpose of training is to determine a set of weights and biases in the neural network that minimizes a loss function. The loss function quantifies the errors between the calculated quantities (energy, force, and virial) from the ML potential and those in the training set, which are usually prepared using quantum mechanical calculations. We denote a set of parameters in the neural network as a vector $\mathbf{z}$, whose dimension is the total number of parameters $N_{\rm par}$. For a neural network with a single hidden layer, 
\begin{equation}
\label{equation:N_par}
N_{\rm par} = (N_{\rm des}+2) N_{\rm neu} + 1. 
\end{equation}
This number is usually a few thousand for the examples studied in this work. We can formally express the loss function as a function of the neural network parameters:
\begin{equation}
L = L(\mathbf{z}),
\end{equation}
and express the training process as a real-valued optimization problem:
\begin{equation}
\mathbf{z}^{\ast} = \min L(\mathbf{z}),
\end{equation}
where $\mathbf{z}^{\ast}$ is an optimal set of parameters. Note that we have used boldface to represent abstract vectors and  italic boldface to represent Cartesian vectors.

The loss function consists of a weighted sum of a few parts. In our formulation, energy, force, and virial can be conveniently calculated, and we thus define loss functions for all of them written as $L_{\rm e}(\mathbf{z})$, $L_{\rm f}(\mathbf{z})$, and $L_{\rm v}(\mathbf{z})$, respectively. In addition, we consider loss functions that serve as regularization terms, which are denoted as $L_{1}(\mathbf{z})$ and $L_{2}(\mathbf{z})$, corresponding to $\mathcal{L}_1$ and $\mathcal{L}_2$ regularizations, respectively. The total loss function is then defined as a weighted sum of all these individual loss functions:
\begin{equation}
\label{equation:loss}
L(\mathbf{z}) = \lambda_{\rm e} L_{\rm e}(\mathbf{z}) +  \lambda_{\rm f} L_{\rm f}(\mathbf{z}) +  \lambda_{\rm v} L_{\rm v}(\mathbf{z}) +  \lambda_1 L_1(\mathbf{z}) +  \lambda_2 L_2(\mathbf{z}).
\end{equation}

The energy loss function is defined as the following root mean square error (RMSE):
\begin{equation}
\label{equation:loss-energy}
L_{\rm e}(\mathbf{z}) 
= \left( 
\frac{1}{N_{\rm str}}\sum_{n=1}^{N_{\rm str}} \left( U^{\rm NEP}(n,\mathbf{z}) - U^{\rm tar}(n)\right)^2
\right)^{1/2},
\end{equation}
where $N_{\rm str}$ is the total number of structures in the training dataset, $U^{\rm tar}(n)$ is the target energy of the $n$th structure, and $U^{\rm NEP}(n,\mathbf{z})$ is the corresponding energy calculated using the NEP potential with the parameters $\mathbf{z}$. Similarly, the force loss function is defined as the following RMSE:
\begin{equation}
\label{equation:loss-force}
L_{\rm f}(\mathbf{z}) 
= \left( 
\frac{1}{3N}
\sum_{i=1}^{N} \left( \bm{F}_i^{\rm NEP}(\mathbf{z}) - \bm{F}_i^{\rm tar}\right)^2
\right)^{1/2},
\end{equation}
where $N$ is the total number of atoms in the training dataset and $\bm{F}_i^{\rm tar}$ and $\bm{F}_i^{\rm NEP}(\mathbf{z})$ are the target force of the $i$th atom and that calculated from the NEP potential with the parameters $\mathbf{z}$, respectively. Lastly, the virial loss function is defined as the following RMSE:
\begin{equation}
\label{equation:loss-virial}
L_{\rm v}(\mathbf{z}) 
= \left( 
\frac{1}{6N_{\rm str}}
\sum_{n=1}^{N_{\rm str}} \sum_{\mu\nu} \left( W_{\mu\nu}^{\rm NEP}(n,\mathbf{z}) - W_{\mu\nu}^{\rm tar}(n)\right)^2
\right)^{1/2},
\end{equation}
where $W_{\mu\nu}^{\rm tar}(n)$ and $W_{\mu\nu}^{\rm NEP}(n,\mathbf{z})$ are the target $\mu\nu$ virial tensor component of the $n$th structure and that calculated from the NEP potential, respectively. 

For the regularization loss functions, we construct them based on the $\mathcal{L}_1$ and $\mathcal{L}_2$ norms of the parameter vector:
\begin{equation}
\label{equation:loss-L1}
L_{1}(\mathbf{z}) 
= \frac{1}{N_{\rm par}} \sum_{n=1}^{N_{\rm par}} |z_n|;
\end{equation}
\begin{equation}
\label{equation:loss-L2}
L_{2}(\mathbf{z}) 
= \left(\frac{1}{N_{\rm par}} \sum_{n=1}^{N_{\rm par}} z_n^2\right)^{1/2}.
\end{equation}
That is, we apply both $\mathcal{L}_1$ and $\mathcal{L}_2$ regularizations to our neural network. The regularization can help to prevent over-fitting by encouraging the weight parameters to develop smaller absolute values than the case without regularization.

The values of the weight parameters in Eq. (\ref{equation:loss}) clearly depend on the units of the relevant quantities. When using eV/atom for energy and virial and eV/\AA~ for force, we find that $\lambda_{\rm e}=\lambda_{\rm f}=\lambda_{\rm v}=1$ is a very good choice. The weight parameters for the regularization terms, $\lambda_{1}$ and $\lambda_{2}$, need to be tuned to keep a good balance between encouraging over-fitting (if $\lambda_{\rm 1}$ and $\lambda_{\rm 2}$ are too small) and under-fitting (if $\lambda_{\rm 1}$ and $\lambda_{\rm 2}$ are too large).

\subsubsection{Separable natural evolution strategy as the training algorithm}

While the gradient descent based back-propagation method is the standard one for training the neural network parameters, here we use the separable natural evolution strategy algorithm \cite{Schaul2011} to train our neural network parameters. This is a principled approach to real-valued evolutionary optimization by following the natural gradient of the loss function to update a search distribution for a population of solutions. It can be considered as a derivative-free blackbox optimizer, which is very suitable to minimize the complex loss function in Eq. (\ref{equation:loss}). 

The workflow of our training algorithm is as follows:
\begin{enumerate}
\item Initialization. Create an initial search distribution in the solution space (of dimension $N_{\rm par}$) with the mean vector $\mathbf{m}$ and standard deviation vector $\mathbf{s}$. The components of $\mathbf{m}$ can be random numbers evenly chosen between $-1/2$ to $1/2$, and the components of $\mathbf{s}$ can be chosen as constants such as $0.1$.
\item Loop over $N_{\rm gen}$ generations:  
\begin{enumerate}
    \item Create a population of solutions $\mathbf{z}_k$ ($1\leq k \leq N_{\rm pop}$) based on the current  $\mathbf{m}$ and $\mathbf{s}$ ($\odot$ denotes elementwise multiplication),
    \begin{equation}
    \mathbf{z}_k \leftarrow \mathbf{m} + \mathbf{s} \odot \mathbf{r}_k,
    \end{equation}
    where $N_{\rm pop}$ is the population size and $\mathbf{r}_k$ is a set of Gaussian-distributed random numbers with mean $0$ and variance $1$. Note that the random number vectors $\mathbf{r}_k$ are different for different individual solutions in the population. 
    \item Evaluate the loss functions $L(\mathbf{z}_k)$ for all the individual solutions $\mathbf{z}_k$ in the population, and sort them  according to  the loss functions, from small to large. 
    \item Update the natural gradients:
    \begin{equation}
        \nabla_{\mathbf{m}} J \leftarrow \sum_{k=1}^{N_{\rm pop}} u_k \mathbf{r}_k;
    \end{equation}
    \begin{equation}
        \nabla_{\mathbf{s}} J \leftarrow \sum_{k=1}^{N_{\rm pop}} u_k (\mathbf{r}_k \odot \mathbf{r}_k-1),
    \end{equation}
    where $u_k$ is a set of rank-based utility values used to evolve the population towards the direction of better individual solutions. For explicit values of $u_k$, see Ref. \onlinecite{Glasmachers2010}.
    \item Update the mean and standard deviation of the search distribution (the exponential function below is applied to its argument in an elementwise way):
    \begin{equation}
        \mathbf{m} \leftarrow \mathbf{m} + \eta_{\mathbf{m}} \left(\mathbf{s} \odot \nabla_{\mathbf{m}} J\right)
    \end{equation}
    \begin{equation}
        \mathbf{s} \leftarrow \mathbf{s} \odot \exp\left(\frac{\eta_{\mathbf{s}}}{2} \nabla_{\mathbf{s}} J \right),
    \end{equation}
    where $\eta_{\mathbf{m}}$ and $\eta_{\mathbf{s}}$ can be understood as the learning rates for $\mathbf{m}$ and $\mathbf{s}$, respectively. We use the suggested values \cite{Schaul2011} of $\eta_{\mathbf{m}}=1$ and $\eta_{\mathbf{s}}=\left(3+\log N_{\rm par}\right)/5\sqrt{N_{\rm par}}$. 
\end{enumerate}
\end{enumerate}

\subsection{Computer implementation\label{section:cuda}}

We have implemented all the relevant calculations (except for the generation of the training data) into the open-source \textsc{gpumd} package \cite{fan2013cpc,fan2017cpc,gpumd} using compute unified device architecture (CUDA). After compiling \textsc{gpumd}, one can obtain three executables, including \verb"nep", \verb"gpumd", and \verb"phonon". The \verb"nep" executable can be used to train a NEP potential, the \verb"gpumd" executable can be used to run MD simulations, and the \verb"phonon" executable can be used to calculate some phonon properties based on harmonic lattice dynamics. All the calculations can be done with the \textsc{gpumd} package without external dependence.  Moreover, both Windows and Linux operating systems are supported, and the  prerequisites for using the \textsc{gpumd} package include only a CUDA-enabled GPU, a suitable CUDA toolkit, and a suitable C++ compiler.

The NEP potential is no more complicated than the empirical Tersoff bond-order potential \cite{tersoff1989prb}, the major difference between them being the different numbers of fitting parameters. Therefore, our GPU implementation of the NEP potential closely follows the implementation of the Tersoff potential \cite{fan2017cpc}. Most importantly, all the per-atom quantities have closed forms in our formulations and we can make a one-to-one correspondence between one atom and one CUDA thread. This is an efficient parallelisation scheme for medium-to-large systems, as it can attain a large degree of parallelism as well as a high arithmetic intensity, both of which are vital for achieving high efficiency in GPU computing. For small systems, this is not an efficient scheme due to the reduced degree of parallelism. Because the systems (structures) in the training dataset are usually quite small (of the order of $100$ atoms), we combine all the individual structures into a single one in the implementation of the NEP potential, following the same strategy as used in the \textsc{gpuga} package \cite{fan2019jpcm} for empirical potential fitting. 

\section{Performance evaluation \label{section:performance}}

In this section, we present a few case studies to evaluate the performance of NEP implemented in \textsc{gpumd}, as compared to the \textsc{quip} \cite{quip} package that implements the GAP-SOAP potential  \cite{bartok2010prl,bartok2013prb}, the \textsc{mlip} package \cite{Novikov2021} that implements the MTP potential \cite{Shapeev2016}, and the DeePMD-kit package \cite{wang2018cpc} that implements the DP potential \cite{zhang2018prl,zhang2018endtoend}. Because a good machine learning potential should be able to account for nearly all the phases of a given material, as demonstrated for elementary silicon \cite{bartok2018prx}, phosphorus \cite{Deringer2020nc}, and carbon \cite{Rowe2020jcp,Muhli2021prb}, we will consider fitting a general-purpose potential for silicon. In addition, we will consider fitting a specific potential for two-dimensional (2D) silicene and a specific potential for bulk PbTe. 

\subsection{Training datasets}

We performed quantum mechanical density functional theory (DFT) calculations to prepare the training data, using \textsc{vasp} \cite{vasp} for bulk PbTe and Quantum Espresso \cite{QE-2009} for 2D silicene. For the general-purpose potential of silicon, we used the training data from Ref. \onlinecite{bartok2018prx}. All the inputs and outputs of the \verb"nep" executable within the \textsc{gpumd} package are openly available in Zenodo \cite{fan2021zenodo}.

\subsubsection{Training dataset for general-purpose silicon} 

The training dataset for general-purpose silicon from Ref. \onlinecite{bartok2018prx} consists of 2475 structures, including an isolated atom providing a reference energy, various three-dimensional (3D) solid structures, sp2 and sp bonded structures, liquid structures, amorphous structures, diamond structures with surfaces, diamond structures with vacancies, and some other defective structures. Every structure has an energy, but not all the structures have virial data. For details on the DFT calculations, the reader is referred to Ref. \onlinecite{bartok2018prx}.

\subsubsection{Training dataset for 2D silicene} 

The training set for 2D silicene consists of $914$ rectangular cells, each with $60$ atoms, obtained via the active learning scheme as implemented in the \textsc{mlip} package \cite{Novikov2021}. We considered states with temperatures ranging from 100 K to 900 K and biaxial in-plane strains ranging from $-1\%$ to $1\%$. For each state, the active learning iteration was terminated when no configurations were selected using the criterion of $\gamma_{\rm select}=2$ \cite{Novikov2021} after 6 independent $10$ ps MD simulations with a pre-trained MTP potential. Static DFT calculations were performed to obtain accurate energy, force, and virial data. For this purpose, we used the PBE functional \cite{pbe}, an optimized norm-conserving Vanderbilt pseudopotential \cite{Schlipf2015cpc}, a kinetic energy cutoff of 40 Ry for wavefunctions, a $3 \times 3 \times 1$ $k$-point mesh, and a threshold of $10^{-10}$ Ry for the electronic self-consistent loop.

\subsubsection{Training dataset for bulk PbTe} 

The training set for bulk PbTe consists of $325$ triclinic cells, each with $250$ atoms. We obtained $305$ cells using DFT-MD simulations with temperatures ranging from $300$ K to $900$ K. We also added $20$ cells with small random atom displacements generated by using the \textsc{hiphive} package \cite{eriksson2019hiphive}, to sample the low-temperature phase space of PbTe. Static DFT calculations were performed to obtain accurate energy and force data, using the PBE functional \cite{pbe}, an energy cutoff of 400 eV for the projector augmented wave \cite{paw1,paw2}, a $1 \times 1 \times 1$ $k$-point mesh, and a threshold of $10^{-7}$ eV for the electronic self-consistent loop.  Virial data were not used for PbTe. This training set has been used to train a GAP-SOAP potential for PbTe \cite{zeng2021prb}.

\subsection{Hyperparameters}

All the machine learning potentials have a number of hyperparameters. We list all the tunable hyperparameters used for the NEP potential in Table \ref{table:nep-hyper}. For the MTP and GAP-SOAP potentials, we present the relevant input commands/scripts in Appendix \ref{appendix:gap_mtp}. For the DP potential, the relevant hyperparameters we used are described in Appendix \ref{appendix:dp}.

\begin{table}[thb]
\centering
\setlength{\tabcolsep}{2Mm}
\caption{The tunable hyperparameters used in the NEP potential for the three materials. Here, $r_{\rm c}^{\rm R}$ ($r_{\rm c}^{\rm A}$) is the cutoff radius for the radial (angular) components of the descriptor, $n_{\rm max}^{\rm R}$ ($n_{\rm max}^{\rm A}$) is the Chebyshev polynomial expansion order for the radial (angular) components, $l_{\rm max}$ is the Legendre polynomial expansion order for the angular components, $N_{\rm neu}$ is the number of neurons in the hidden layer of the neural network, $\lambda_1$ ($\lambda_2$) is the $\mathcal{L}_1$ ($\mathcal{L}_2$) regularization parameter, $N_{\rm pop}$ is the population size in the natural evolution strategy algorithm, and $N_{\rm gen}$ is the maximum number of generations to be evolved.}
\label{table:nep-hyper}
\begin{tabular}{llll}
\hline
\hline
Parameter & general Si & Silicene & PbTe  \\
\hline
$r_{\rm c}^{\rm R}$  & 5 \AA & 5.5 \AA & 8 \AA  \\
$r_{\rm c}^{\rm A}$  & 5 \AA & 5.5 \AA & 4 \AA  \\
$n_{\rm max}^{\rm R}$  & $15$ & $12$ & $12$  \\
$n_{\rm max}^{\rm A}$  & $15$ & $12$ & $6$  \\
$l_{\rm max}$  & $4$  & $4$ & $4$  \\
$N_{\rm neu}$  & $50$ & $40$ & $40$ \\
$\lambda_{1}$  & $0.05$ & $0.05$ & $0.05$ \\
$\lambda_{2}$  & $0.05$ & $0.05$ & $0.05$ \\
$N_{\rm pop}$  & $50$ & $50$ & $50$ \\
$N_{\rm gen}$  & $2\times 10^5$ & $2\times10^5$ & $2\times 10^5$ \\
\hline
\hline
\end{tabular}
\end{table}

\begin{figure}[htb]
\begin{center}
\includegraphics[width=\columnwidth]{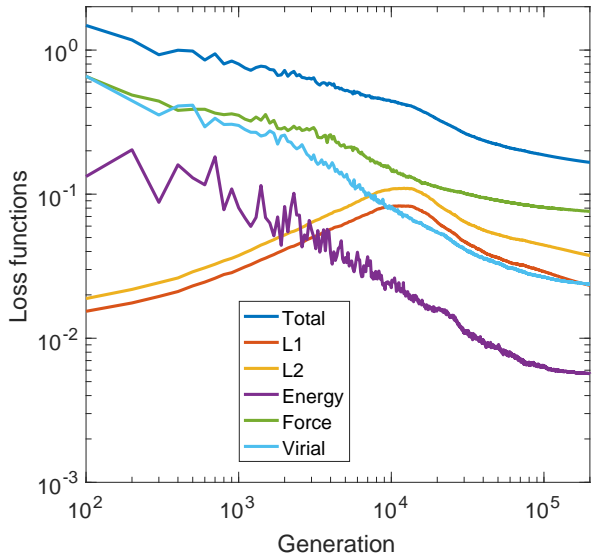}
\caption{Evolution of the loss functions during the training process. ``Total'', ``L1'', ``L2'', ``Energy'', ``Force'',  and ``Virial'' correspond to Eq. (\ref{equation:loss}), Eq. (\ref{equation:loss-L1}), Eq. (\ref{equation:loss-L2}), Eq. (\ref{equation:loss-energy}), Eq. (\ref{equation:loss-force}), and Eq. (\ref{equation:loss-virial}), respectively. The training set for general silicon systems \cite{bartok2018prx} is used here and the hyperparameters are presented in Table \ref{table:nep-hyper}. }
\label{figure:loss}
\end{center}
\end{figure}

For general silicon and 2D silicene, which mainly have covalent bonding, we find it not beneficial to use different $r_{\rm c}^{\rm R}$ and $r_{\rm c}^{\rm A}$. Accordingly, $n_{\rm max}^{\rm R}$ and $n_{\rm max}^{\rm A}$ are chosen to be the same. However, for PbTe, which is an ionic crystal having relatively long-ranged (Coulomb) interactions, we find it beneficial to use larger $r_{\rm c}^{\rm R}$ and $n_{\rm max}^{\rm R}$ for the radial components of the descriptor, and use smaller $r_{\rm c}^{\rm A}$ and $n_{\rm max}^{\rm A}$ for the angular components, which can reduce the computational cost with a given target accuracy.

For all the materials, we only use a single hidden layer with $40$ or $50$ neurons in the neural network, which is sufficient based on our tests. We note that the deep neural network potential  \cite{wang2018cpc,zhang2018prl} usually requires using a deep neural network with several hidden layers, each with a large number of neurons. This is because a relatively simple (but very general) atom environment descriptor was used in the deep neural network potential \cite{wang2018cpc,zhang2018prl} and it requires to use a deep neural network to establish a connection between the simple descriptor and the energy of an atom. 

For the regularization parameters, we find that $\lambda_1=\lambda_2=0.05$ is a good default choice. The remaining two parameters, $N_{\rm pop}$ and $N_{\rm gen}$, are only relevant for the training process. The computation time in the training process is proportional to the product of them. 

The dimension of the descriptor $N_{\rm des}$ for each training set can be calculated according to Eq. (\ref{equation:N_des}). It is $80$, $65$ and $41$ for general silicon, 2D silicence and bulk PbTe, respectively. Therefore, the structures of the neural networks for these training sets can be denoted as $80$-$50$-$1$, $65$-$40$-$1$, and $41$-$40$-$1$, respectively. The numbers of fitting parameters $N_{\rm par}$ in these neural networks are respectively $4101$, $2681$, and $1721$, according to Eq. (\ref{equation:N_par}).

Figure \ref{figure:loss} shows the evolution of the various loss functions with respect to the generation, for the case of general silicon. With increasing generation, the RMSEs of energy, force, and virial are reduced and converged, although with some oscillations in the beginning. In contrast, the loss functions for the $\mathcal{L}_1$ and $\mathcal{L}_2$ regularization, hence the magnitudes of the weight and bias parameters in the neural network, first increase and then decrease, indicating the effectiveness of the regularization. Without the regularization (that is, setting $\lambda_1$ and $\lambda_2$ to zero or very small values), the weight and bias parameters in the neural network would increase wildly, which can easily lead to over-fitting. Explicit examples of over-fitting and under-fitting are demonstrated in Appendix \ref{appendix:regularization}.

We note that the training process is very stable: independent runs with different sets of random numbers lead to comparable results with very small variation. This strongly suggests that the natural evolution strategy we used can find very good minima of the loss function with sets of globally optimized neural network parameters.

\subsection{Evaluation of the accuracy}

\begin{table}[thb]
\centering
\setlength{\tabcolsep}{2Mm}
\caption{Accuracy comparison between NEP, GAP-SOAP, MTP, and DP. Energy and virial RMSEs are in units of meV/atom, and force RMSE is in units of meV/\AA. }
\label{table:rmse_silicene_pbte}
\begin{tabular}{llllll}
\hline
\hline
Material & RMSE  & GAP & MTP & DP & NEP\\
\hline
\multirow{3} {*} {Silicene} 
& Energy    & $1.6$    & $1.3$ & $1.8$  & $1.5$ \\
& Force    & $65$      & $50$  & $65$   & $56$ \\
& Virial    & $14$    & $10$  & $8.1$ & $8.8$ \\
\hline
\multirow{2} {*} { PbTe}   
& Energy    & $0.50$ & $0.63$ & $0.63$  & $0.56$\\
& Force     & $50$            & $52$  & $53$   & $50$\\
\hline
\hline
\end{tabular}
\end{table}

\begin{figure}[htb]
\begin{center}
\includegraphics[width=\columnwidth]{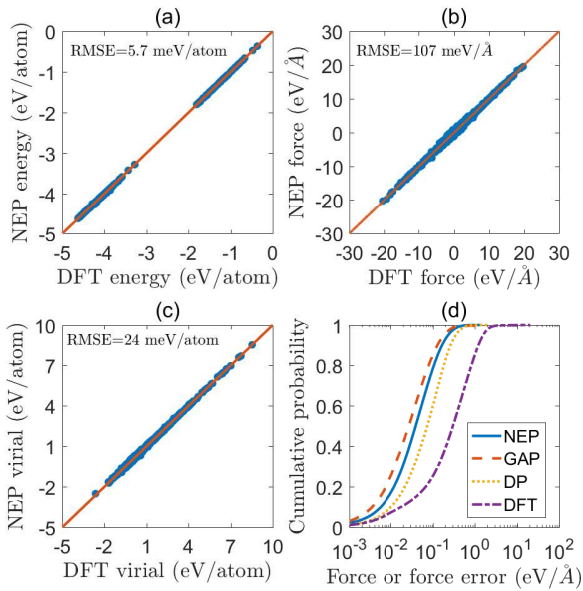}
\caption{(a) Energy, (b) force, and (c) virial as calculated from the NEP potential compared with the training data from quantum mechanical DFT calculations. The solid lines in (a)-(c) represent the identity function used to guide the eyes. (d) Cumulative probability as a function of the force component error from NEP (solid line), GAP-SOAP (dashed line), and DP (dotted line). The cumulative probability of the absolute DFT force components (dot-dashed line) is shown as a reference.}
\label{figure:error}
\end{center}
\end{figure}

Figure \ref{figure:error} compares the predicted energy, force and virial values by NEP and those from quantum mechanical DFT calculations for the general silicon training set \cite{bartok2018prx}. The agreement between NEP and DFT is reasonably good. Based on these data, we calculate the RMSEs, as given in panels (a) to (c). Reference \onlinecite{bartok2018prx} did not provide these RMSE values. Instead, the cumulative probability of the force component errors between GAP-SOAP and DFT calculations was provided. We therefore compare this quantity from NEP and GAP-SOAP in Fig. \ref{figure:error}(d). It can been seen that NEP is slightly less accurate than GAP-SOAP here. Figure \ref{figure:error}(d) also shows that DP (force RMSE is $165$ meV/\AA) is less accurate than NEP (force RMSE is $107$ meV/\AA) in this case.

We similarly calculated the RMSEs for 2D silicene and bulk PbTe and the values are listed in Table \ref{table:rmse_silicene_pbte}. It can bee seen that all the potentials have comparable accuracy.
A given ML potential is not always more accurate than another one, as the accuracy of a potential is determined by many tunable hyperparameters. The hyperpameters affect not only the accuracy, but also the computational efficiency, an important issue that we will discuss next.

\begin{figure}[htb]
\begin{center}
\includegraphics[width=\columnwidth]{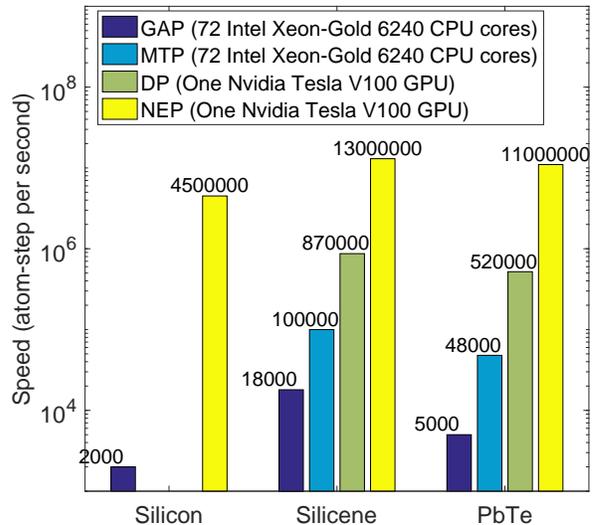}
\caption{Computational speed of the NEP potential (running with an Nvidia Tesla V100 GPU), compared to those of the DP potential (running with the same V100 GPU) and the GAP-SOAP and MTP potentials (both running with $72$ Intel Xeon-Gold $6240$ CPU cores). The model compression technique \cite{lu2021dp} is applied to the DP potentials here to speed up the calculations.}
\label{figure:speed}
\end{center}
\end{figure}

\subsection{Evaluation of the computational efficiency}

\subsubsection{Computational speed}

In Fig. \ref{figure:speed}, we compare the computational speeds of the four ML potentials in MD simulations. The computational speed is measured as the product of the number of atoms and the number of MD steps divided by the computation time, which is in units of atom-step per second. To measure the speed, we used a simulation cell with $8\times 10^3$ (for bulk silicon and PbTe) or $6\times 10^3$ (for 2D silicene) atoms and run an MD simulation for $100$ steps in the NVT ensemble, outputting basic thermodynamic quantities every $10$ steps. The \textsc{quip} \cite{quip}, \textsc{mlip} \cite{Novikov2021}, and DeePMD-kit \cite{wang2018cpc} packages were built as libraries to be called by the \textsc{lammps} package \cite{plimpton1995jcp}. 

Both the GAP-SOAP and MTP potentials are implemented in CPU only, and we used $72$ Intel Xeon-Gold $6240$ CPU cores (two nodes, each with $36$ cores) parallelized by the message passing interface (MPI). Note that the GAP-SOAP potential achieves very high accuracy for general silicon, but has very low computational speed. Reference \onlinecite{bartok2018prx} used $n_{\rm max}=10$ and $l_{\rm max}=12$ in the SOAP descriptor, together with $9\times 10^3$ basis functions in the Gaussian process regression. Using smaller values of these parameters will increase the computational speed but reduce the accuracy. With comparable accuracy, MTP is about one order of magnitude faster than GAP-SOAP, as has been reported before \cite{Shapeev2016}.

Our NEP potential is implemented in GPU only, and we used an Nvidia Tesla V100 GPU card in the performance test. Note that the above CPU and GPU resources are of comparable price. The computational speed of \textsc{gpumd} can be above $10^7$ atom-step per second, which is only about one order of magnitude lower than the computational speed of efficient empirical potentials such as the Tersoff potential as implemented in \textsc{gpumd}. To achieve the same computational speed of \textsc{gpumd} with one V100 GPU, one needs to run \textsc{quip} with at least a few tens of thousand CPU cores, or run \textsc{mlip} with at least a few thousand CPU cores, even assuming an ideal MPI scaling. 

The DeePMD-kit package \cite{wang2018cpc} has both CPU and GPU versions and we used the latter with the same V100 GPU. We used the recent model compression technique \cite{lu2021dp} introduced to the DP potential to speed up the calculations. From Fig. \ref{figure:speed} we see that NEP as implemented in \textsc{gpumd} is more than one order of magnitude faster than DP as implemented in DeePMD-kit, using the same computational resource. 

\subsubsection{Memory consumption}

Apart from the high computational speed, our implementation of NEP in \textsc{gpumd} is also memory efficient. The major memory consumption in the NEP potential is as follows: three neighbor lists (each with a different cutoff radius; we use multiple levels of neighbor list to save computations) which require about $12NM$ bytes, the partial forces $\{\partial U_i/\partial \bm{r}_{ij}\}$ for all atom pairs within a cutoff radius which require about $24NM$ bytes, and some intermediate results related to the descriptor which require about $2000N$ bytes. Here, $N$ is the number of atoms and $M$ is an estimated upper bound of the number of neighbors that an atom can ever have in a given application. Assuming a typical value of $M=100$, the total amount of memory listed above is less than six gigabytes for a one-million-atom system. Therefore, using a single V100 GPU with $32$ gigabytes of device memory, one can simulate systems with up to a few million atoms using the NEP potential in \textsc{gpumd}.

\section{Heat transport applications\label{section:heat}}

MD simulations with ML potentials have been applied to study heat transport properties of a number of materials, including, e.g., GeTe and MnGe compounds \cite{Sosso2012prb,Campi2015jap,bosoni2019jpd,Mangold2020jap}, diamond and amorphous silicon \cite{Huang2019prb,qian2019mtp,li2020mtp}, multilayer graphene \cite{wen2019prb}, monolayer silicene \cite{zhang2019jap}, CoSb$_3$ \cite{Korotaev2019PRB}, monolayer MoS$_2$ and MoSe$_2$ and their alloys \cite{gu2019cms}, C$_3$N \cite{Mortazavi_2020}, $\alpha$-Ag$_2$Se \cite{shimamura2020jcp,shimamura2021cpl}, $\beta$-Ga$_2$O$_3$ \cite{li2020apl}, Tl$_3$VSe$_4$ \cite{zeng2021prb}, PbTe \cite{zeng2021prb}, and SnSe \cite{liu2021jpcm}. There are also works that exclusively used the Boaltzmann transport equation (BTE) approach to calculate thermal conductivity based on force constants determined from ML potentials \cite{Minamitani2019ape,babaei2019prm,Rodriguez2020prb,George2020jcp,Liu2020jcp,Mortazavi2021cpc}. In this section, we use 2D silicene and bulk PbTe as the examples to demonstrate the applicability of NEP in heat transport calculations.

\begin{figure}[htb]
\begin{center}
\includegraphics[width=\columnwidth]{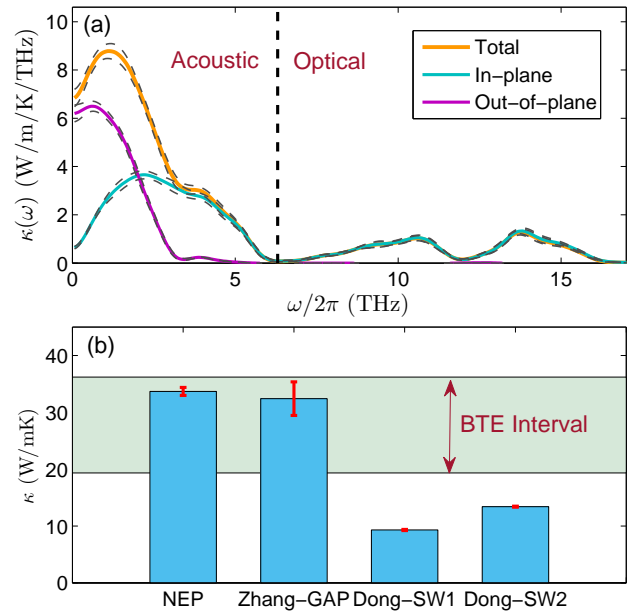}
\caption{(a) Spectral lattice thermal conductivity of 2D silicene at $300$ K calculated using the NEP potential. Solid lines are the average over 10 independent runs and dashed lines represent the error bounds. The vertical dashed line separates the acoustic and optical phonon frequencies. (b) Lattice thermal conductivity of 2D silicene at 300 K calculated using MD simulations with the NEP potential, MD simulations with a GAP-SOAP potential \cite{zhang2019jap}, and MD simulations with two versions of the Stillinger-Weber potential \cite{dong2018prb}. The area between the two horizontal lines represents the range of values predicted by BTE-DFT calculations.}
\label{figure:kappa_sil}
\end{center}
\end{figure}

\begin{figure}[htb]
\begin{center}
\includegraphics[width=\columnwidth]{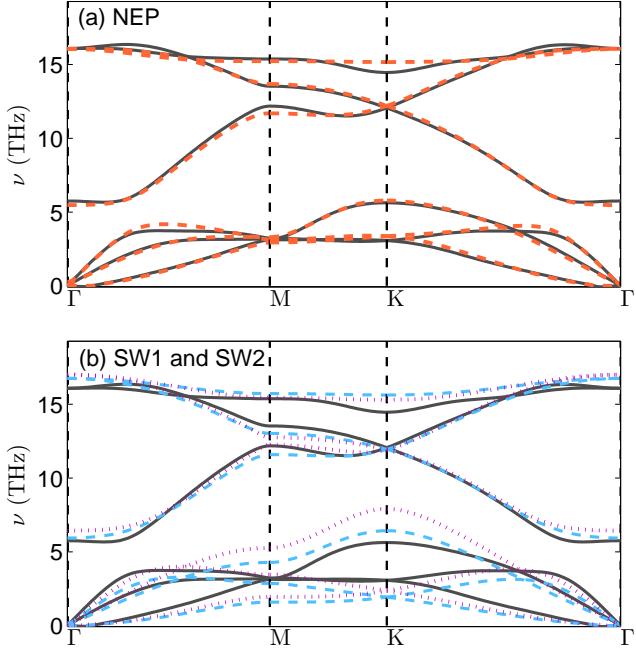}
\caption{Phonon dispersions of silicene. (a) Comparison between NEP (dashed line) and DFT (solid line) calculations. (b) Comparison between SW1 (dashed line), SW2 (dot-dashed line) and DFT (solid line) calculations. The SW1 and SW2 results are from Ref. \onlinecite{zhang2019jap}. }
\label{figure:phonon_sil}
\end{center}
\end{figure}

We calculate the lattice thermal conductivity using the HNEMD method and the related spectral decomposition method \cite{Fan2019prb} as implemented in \textsc{gpumd} and reviewed in Sec. \ref{section:virial-heat}. The simulation cells for 2D silicene and bulk PbTe contain $6000$ and $8000$ atoms, respectively, which are large enough to eliminate finite-size effects. For silicene, we performed $10$ independent runs, each with a $50$ ps equilibration stage in the NPT ensemble (with a target in-plane pressure of zero) and a $1000$ ps production stage with an external driving force of $0.5$ $\mu$m$^{-1}$. For PbTe, we performed $3$ independent runs, each with a $100$ ps equilibration stage in the NVT ensemble and a $2000$ ps production stage with an external driving force of $1.0$ $\mu$m$^{-1}$. The integration time step is $0.5$ fs for silicene and $1.0$ fs for PbTe. We consider a temperature of $300$ K for silicene and a temperature range from $300$ to $700$ K for PbTe.

\subsection{Thermal transport in 2D silicene}

Figure \ref{figure:kappa_sil}(a) shows the spectral lattice thermal conductivity of silicene for the in-plane and out-of-plane phonons (also called flexural phonons), according to the spatial decomposition \cite{fan2017prb} for 2D materials. The out-of-plane phonons only have nonzero lattice thermal conductivity for the acoustic branches, while the in-plane phonons have nonzero lattice thermal conductivity for both the acoustic and optical branches. In total, the lattice thermal conductivity is mainly contributed by the acoustic phonons. 

Summing up the various components in Fig. \ref{figure:kappa_sil}(a), we obtain a total lattice thermal conductivity of $33.7\pm 0.6$ W/mK. Our value is consistent with that calculated \cite{zhang2019jap} using the GAP-SOAP potential ($32.4\pm 2.9$ W/mK), as shown in Fig. \ref{figure:kappa_sil}(b). Our lattice thermal conductivity value is also well within the range from previous BTE-DFT calculations \cite{gu2015jap,kuang2016nanoscale,xie2016prb,peng2016prb}. In contrast, the lattice thermal conductivity values from MD simulations \cite{dong2018prb} using two versions (called SW1 and SW2) of the Stillinger-Weber potential \cite{stillinger1985prb} optimized for silicene \cite{zhang2014prb} are well below this range. The underestimation of the lattice thermal conductivity by the Stillinger-Weber potential can be partially understood by examining the phonon dispersions, as shown in Fig.  \ref{figure:phonon_sil}. Both versions of the Stillinger-Weber potential significantly underestimate the frequencies and hence the group velocities of the phonons around the $\Gamma$ point as compared to the DFT results. On the other hand, the phonon dispersions from the NEP potential agree well with the DFT results.  

\subsection{Thermal transport in bulk PbTe}

\begin{figure}[htb]
\begin{center}
\includegraphics[width=\columnwidth]{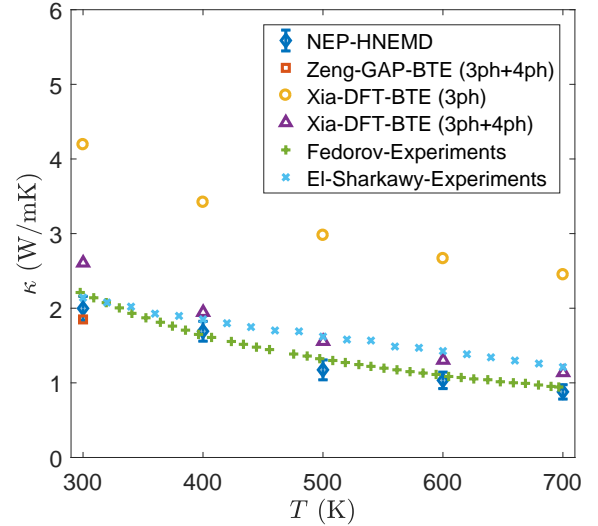}
\caption{Lattice thermal conductivity of bulk PbTe as a function of temperature, from HNEMD simulations with the NEP potential, from BTE calculations with the GAP-SOAP potential considering both three-phonon and four-phonon scatterings \cite{zeng2021prb}, from BTE-DFT calculations considering three-phonon scattering only or both three-phonon and four-phonon scatterings \cite{xia2018apl}, and from  experiments \cite{Fedorov1969,El-Sharkawy1983ijt}.}
\label{figure:kappa_pbte}
\end{center}
\end{figure}

We similarly calculated the thermal conductivity of PbTe from $300$ to $700$ K, and the results are shown in Fig. \ref{figure:kappa_pbte}. The predictions by NEP agree well with the experimental data \cite{Fedorov1969,El-Sharkawy1983ijt}. BTE calculations with force constants from DFT \cite{xia2018apl} or the GAP-SOAP potential \cite{zeng2021prb} considering both three-phonon and four-phonon scatterings also produce comparable results. However, if only three-phonon scattering is considered in the BTE calculations \cite{xia2018apl}, the obtained lattice thermal conductivity values are about two times as large. This indicates the importance of four-phonon scattering in PbTe. In some other materials such as Tl$_3$VSe$_4$, it has been suggested that the perturbation theory as adopted in the BTE approach can severely underestimate the phonon scatterings even when four-phonon scattering is included \cite{zeng2021prb}. Because phonon anharmonicity is fully taken into account in MD simulations, we expect that our efficient NEP potential will serve as a promising tool for investigating heat transport properties in materials with strong phonon anharmonicity. 

\subsection{Thermal transport in amorphous silicon}

Our NEP potential can also be applied to study heat transport in materials with strong spatial disorder. We show this by considering amorphous silicon (a-Si), using the general silicon NEP potential fitted above. To generate a-Si structures, we follow the MD simulation protocol as described in Ref. \onlinecite{bartok2018prx}. Because our NEP potential is much more efficient than the GAP-SOAP potential, we use a simulation cell with $64000$ atoms instead of $64$ atoms as used in Ref. \onlinecite{bartok2018prx}.

\begin{figure}[htb]
\begin{center}
\includegraphics[width=\columnwidth]{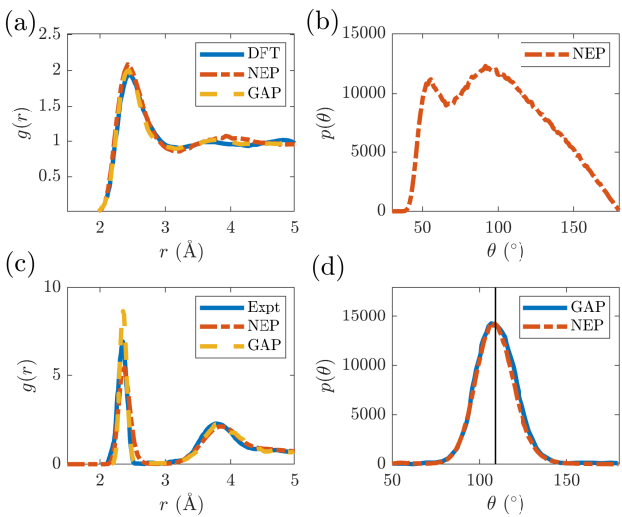}
\caption{(a) Radial and (b) angular distribution functions of liquid silicon ($2000$ K). (c) Radial and (d) angular distribution functions of a-Si ($500$ K). DFT and GAP results in (a-c) are from Ref. \onlinecite{bartok2018prx}. GAP results in (d) are from Ref. \onlinecite{Deringer2018jpcl}. The vertical solid line in (d) represents the bond angle (about $109.5^{\circ}$) in the diamond structure.}
\label{figure:a-si-structure}
\end{center}
\end{figure}

\begin{figure}[htb]
\begin{center}
\includegraphics[width=\columnwidth]{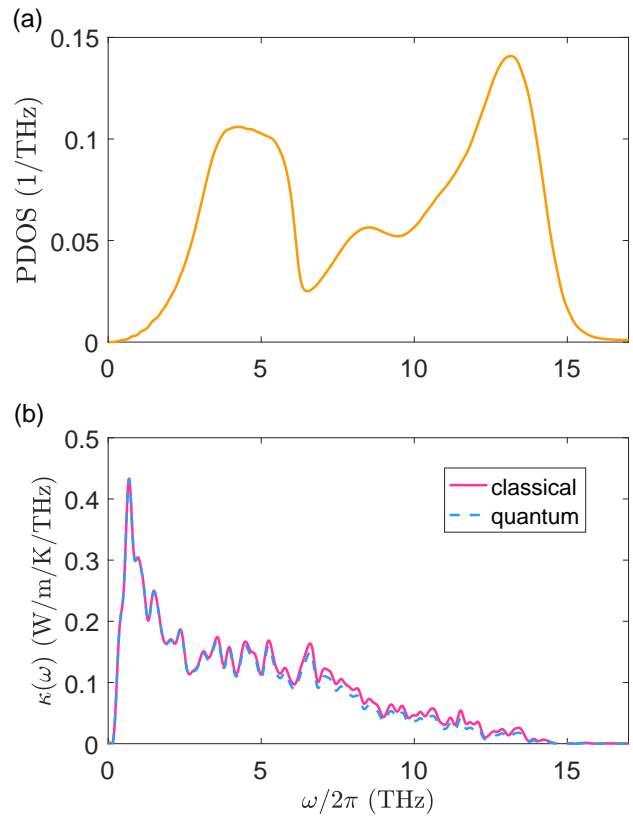}
\caption{(a) PDOS and (b) spectral thermal conductivity of a-Si at $300$ K and zero pressure. In (b), the solid line represents the results from classical MD simulations and the dashed line represents the quantum-corrected results. See text for details. }
\label{figure:a-Si-thermal}
\end{center}
\end{figure}

The structural properties during the process of generating a-Si are shown in Fig. \ref{figure:a-si-structure}. The radial distribution function $g(r)$ and the angular distribution function $p(\theta)$ at liquid ($2000$ K) and amorphous ($500$ K) states from NEP agree well with the DFT and GAP results. This ensures that the a-Si ``samples'' we constructed have reasonable structural properties.

Figure \ref{figure:a-Si-thermal} shows the calculated phonon density of states (PDOS) from velocity-autocorrelation function and spectral lattice thermal conductivity from virial-velocity correlation function of a-Si at $300$ K and zero pressure, averaged over five a-Si samples. Although the high-frequency phonons have a large contribution to the PDOS, the thermal conductivity is still mainly contributed by low-frequency phonons, similar to the case of crystalline 2D silicene above. The total lattice thermal conductivity of our a-Si samples is calculated to be $1.5 \pm 0.1$ W/mK, from the classical MD simulations. Although there is so far no universally applicable quantum correction method for classical MD simulations of phonon thermal transport in perfect crystals, a feasible one \cite{lv2016njp,saaskilahti2016aipadvance,fan2017nl} for disordered materials is to multiply a factor to the classical spectral thermal conductivity $\kappa_{\mu\nu}(\omega)$ to obtain the following quantum-corrected spectral thermal conductivity:
\begin{equation}
\kappa^{\rm q}_{\mu\nu}(\omega) = \kappa_{\mu\nu}(\omega) \frac{x^2e^x}{(e^x-1)^2}.
\end{equation}
Here, $x=\hbar\omega/k_{\rm B}T$, where $\hbar$ and $k_{\rm B}$ are the reduced Planck constant and the Boltzmann constant, respectively. The quantum-corrected spectral thermal conductivity is also shown in Fig. \ref{figure:a-Si-thermal}. We see that at $300$ K, the quantum effects are very minor, which only reduce the total thermal conductivity to $1.4 \pm 0.1$ W/mK. However, we note that the quantum effects can be very strong at low temperatures. Our results here agree well with previous studies using the GAP-SOAP potential \cite{qian2019mtp} and the DP potential \cite{li2020mtp}. A more  systematical study considering different temperatures, strains (stresses), cell sizes, and quenching rates is beyond the scope of this paper.

\section{Summary and conclusions \label{section:summary}}

In summary, we have presented NEP, a framework for generating neural network based ML potentials using the separable natural evolutionary strategy algorithm and an atom-environment descriptor based on Chebyshev and Legendre polynomials. Our NEP potential can achieve accuracy comparable to that of the other ML potentials, such as GAP-SOAP, MTP, and DP, and our GPU implementation attains a high computational efficiency, in terms of both computation time and memory usage. The computation time of the NEP potential is of the order of $0.1$ microsecond per atom per MD step using a single Nvidia V100 GPU, and one can use this amount of resource to simulate systems up to a few million atoms.

Combining with the HNEMD and the related spectral decomposition methods, our work makes it possible to efficiently and accurately simulate heat transport in various materials based on quantum-mechanical training data only. We expect that the NEP approach will be particularly useful for modeling heat transport properties of materials with strong phonon anharmonicity or spatial disorder, which usually cannot be accurately treated either with traditional empirical potentials or with the perturbative BTE method.

\begin{acknowledgments}
ZF acknowledges the supports from the National Natural Science Foundation of China (NSFC) (No. 11974059). ZZ and YC are grateful for the research computing facilities offered by ITS, HKU. YW, HD, and TA-N acknowledge the supports from the Academy of Finland Centre of Excellence program QTF (Project 312298) and the computational resources provided by Aalto Science-IT project and Finland's IT Center for Science (CSC).
\end{acknowledgments}

\appendix

\section{The Chebyshev polynomials used in NEP\label{appendix:chebyshev}}

The Chebyshev polynomials of the first kind are defined in terms of the initial values $T_0(x)=1$ and $T_1(x)=x$, and the recurrence relation ($n \geq 2$)
\begin{equation}
    T_n(x) = 2 x  T_{n-1}(x) - T_{n-2}(x).
\end{equation}
The derivative $dT_n(x)/dx$ is related to the Chebyshev polynomials of the second kind $U_{n-1}(x)$ for $n>0$:
\begin{equation}
    \frac{dT_n(x)}{dx} = n U_{n-1}(x).
\end{equation}
The Chebyshev polynomials of the second kind are defined in terms of the initial values $U_0(x)=1$ and $U_1(x)=2x$, and the recurrence relation ($n \geq 2$)
\begin{equation}
    U_n(x) = 2 x  U_{n-1}(x) - U_{n-2}(x).
\end{equation}

\section{Inputs for training the MTP and GAP potentials\label{appendix:gap_mtp}}

In this Appendix, we give the input commands/scripts used for training the MTP and GAP potentials. 

\subsection{Inputs for training the MTP potential of 2D silicene}

\begin{verbatim}
MTP
version = 1.1.0
potential_name = MTP1m
species_count = 1
potential_tag =
radial_basis_type = RBChebyshev
        min_dist = 1
        max_dist = 5.5
        radial_basis_size = 8
        radial_funcs_count = 5
        alpha_moments_count = 1352
alpha_index_basic_count = 295
\end{verbatim}

\subsection{Inputs for training the MTP potential of bulk PbTe}
\begin{verbatim}
MTP
version = 1.1.0
potential_name = MTP1m
species_count = 2
potential_tag =
radial_basis_type = RBChebyshev
        min_dist = 2.6
        max_dist = 6.0
        radial_basis_size = 8
        alpha_moments_count = 718
alpha_index_basic_count = 201
\end{verbatim}

\subsection{Inputs for training the GAP potential of 2D silicene}

\begin{verbatim}
gap_fit atoms_filename=train.xyz 
gap={distance_2b cutoff=5.5 n_sparse=50
covariance_type=ard_se delta=2.0
theta_uniform=1.0 sparse_method=uniform :
angle_3b cutoff=3.0 n_sparse=120
covariance_type=ard_se delta=1.0
theta_uniform=1.0 sparse_method=uniform : 
soap l_max=6 n_max=12 atom_sigma=0.5 zeta=4
cutoff=5.5 cutoff_transition_width=0.5
n_sparse=600 delta=0.5
covariance_type=dot_product
sparse_method=cur_points} 
default_sigma={0.001 0.001 0.001 0} 
sparse_jitter=1.0e-10
hessian_parameter_name=dummy
virial_parameter_name=virial
energy_parameter_name=energy
force_parameter_name=force
\end{verbatim}

\subsection{Inputs for training the GAP potential of bulk PbTe}

\begin{verbatim}
gap_fit atoms_filename=train.xyz
default_sigma={0.001 0.04 0.04 0}
gap={distance_2b cutoff=6
covariance_type=ard_se delta=0.5
theta_uniform=1.0 sparse_method=uniform
add_species=T n_sparse=20: 
soap l_max=6 n_max=12 cutoff=6
cutoff_transition_width=1.0 delta=1.0
atom_sigma=0.5 zeta=4 
sparse_method=cur_points add_species=T
n_sparse=600 covariance_type=dot_product}
energy_parameter_name=energy
force_parameter_name=forces
\end{verbatim}

\section{Hyperparameters used in the DP potentials\label{appendix:dp}}

For the DP potential, we trained the smooth edition \cite{zhang2018endtoend} for general silicon, 2D silicene and bulk PbTe, using the DeePMD-kit package \cite{wang2018cpc}. The cutoff distance is $5$ \AA~ for general silicon, $5.5$ \AA~ for silicene, and $6$ \AA~ for PbTe. The number of training steps is $4\times10^6$ for general silicon and $10^6$ for both silicene and PbTe. Below are the common hyperparameters used for all the materials. The smoothing parameter is $2$ \AA. The size of the embedding net is $(25,50,100)$, and the size of the fitting net is $(240,240,240)$. The learning rate in the stochastic gradient descent algorithm decreases exponentially from $10^{-3}$ to $10^{-8}$. The weighting parameters for energy, force, and virial (if exists) have a starting value of $0.02$, $1000$, and $1$, respectively, which are linearly changed to $1$ during the training process. The maximum number of neighbors is estimated to be $100$. A default batch-size of $1$ is used.

\section{Effects of regularization\label{appendix:regularization}}

We use the case of 2D silicene to demonstrate the effects of the regularization. We first train a NEP potential with $714$ structures (the training set), and then make predictions for another $200$ structures (the testing set). The testing set contains structures with pressures and temperatures that are not covered by the training set. Therefore, evaluating the trained NEP potential against the testing set involves both interpolation and extrapolation, which serves a good purpose of detecting possible over-fitting and under-fitting.  

\begin{figure}[htb]
\begin{center}
\includegraphics[width=\columnwidth]{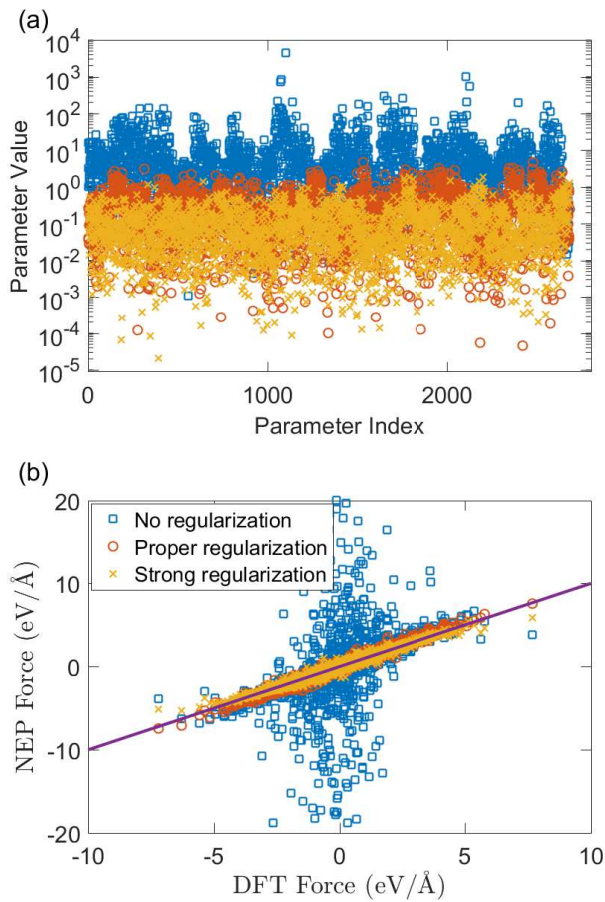}
\caption{(a) The absolute values of the $2681$ neural network parameters in the NEP potential for 2D silicene trained with different levels of regularization. (b) The corresponding predicted forces versus the DFT reference values. }
\label{figure:reg}
\end{center}
\end{figure}

Figure \ref{figure:reg}(a) shows the absolute neural network parameters in the NEP potentials with three levels of regularization: $\lambda_1=\lambda_2=0$ (no regularization), $\lambda_1=\lambda_2=0.05$ (proper regularization), and $\lambda_1=\lambda_2=0.5$ (strong regularization). We see that the average magnitude of the parameters decreases with increasing level of regularization. Quantitatively, the mean absolute values of the parameters are $12$, $0.36$, and $0.12$, respectively, for the three regularization levels above. Figure \ref{figure:reg}(b) clearly shows that the potential trained without regularization is over-fitted, which has very large errors in the predicted forces. On the other hand, the potential trained with too strong a regularization is under-fitted and also has significant errors in the predicted forces. By contrast, the potential trained with proper regularization has the highest accuracy in the testing set, showing the capability of interpolating and even extrapolating (to some extent).

\end{document}